\documentclass[useAMS,usenatbib]{mn2e}
\usepackage{graphicx}
\title[Chromospheric Thermal Continuum Millimetre Emission from non-dusty K and M Red Giants]{Chromospheric Thermal Continuum Millimetre Emission from non-dusty K and M Red Giants}
\author[G. M. Harper, N. O'Riain and T. R. Ayres]{G. M. Harper$^1${\thanks{E-mail:
graham.harper@tcd.ie}, N. O'Riain$^1$ and T. R. Ayres$^2$}\\
$^1$School of Physics, Trinity College, Dublin 2, Ireland \\
$^2$Center for Astrophysics and Space Astronomy, University of Colorado, Boulder, CO, USA}
\begin{document}

\date{Submitted June 12th 2012}

\pagerange{\pageref{firstpage}--\pageref{lastpage}} \pubyear{2012}

\maketitle

\label{firstpage}

\begin{abstract}
We examine the thermal free-free millimetre fluxes expected from non-dusty and non-pulsating 
K through mid-M giant stars based on our limited understanding of their inhomogeneous chromospheres. 
We present a semi-analytic model that provides estimates of the radio fluxes for the  mm wavelengths
(e.g., CARMA, ALMA, JVLA Q-band) based on knowledge of the effective temperatures, angular diameters and
chromospheric Mg~II h \& k emission fluxes. At 250~GHz, the chromospheric optical depths are expected to be
significantly less than unity, which means that fluxes across the mm and sub-mm range will have a contribution from the chromospheric material that gives rise to the ultraviolet emission spectrum, as well as the
cool molecular material known to exist above the photosphere. We predict a lower bound to the inferred brightness temperature of red giants based on heating at the {\it basal-flux limit} if the upper chromospheres have 
filling factor $\simeq 1$. Multi-frequency  mm observations should provide important new information on the structuring of the inhomogeneous chromospheres, including the boundary layer, and allow tests of competing theoretical models for atmospheric heating. We comment on the suitability of these stars as mm flux 
calibrators.
\end{abstract}

\section{Introduction}

Convective motions beneath the photospheres of late-type (cool) stars agitate
the plasma, leading to the excitation of
acoustic and magnetic disturbances. These waves induce non-radiative heating of the upper photosphere and chromosphere. Essentially all cool stars possess material heated above the
prediction of radiative equilibrium, which emits a chromospheric
spectrum that is most conspicuous in the ultraviolet (uv). The 
specific roles of acoustic and magnetic waves in heating the quiet solar chromosphere 
and the role of magnetic fields in structuring its atmosphere are the subject of
considerable debate \citep*{kalkofen_etal1999,ayres2002,vecchio_etal2009}. 
For the Sun and red giants the
chromospheric plasma has a detectable signature in the millimetre (mm)
thermal continuum \citep*{altenhoff_etal1994,loukitcheva_etal2004}. 

Stellar chromospheric surface fluxes measured from 
dominant emission lines, such as Ca~II H \& K and Mg~II h \& k, show a wide scatter at a given effective temperature ($T_{\rm eff}$), 
just as found for regions of different activity on the Sun.
However, there is a well defined empirical lower bound known as the {\it basal-flux},
which is a strong function of effective temperature, $F_{\rm basal} \sim T_{\rm eff}^7$, but is 
only a weak function of surface gravity ($g_\ast$). The chromospheric cooling from the quietest solar regions, in the centres of supergranules, is close to the {\it basal-flux} for a $T_{\rm eff}=5770$~K star (e.g., \citealt{schrijver_zwaan2000}).
 Stars that have chromospheric emission above the {\it basal-flux} are thought to have additional
contributions from areas of enhanced surface magnetic activity, analogous to solar active regions (sunspots and associated plage). However, no consensus has been reached
about the origin of the {\it basal-flux}. One line of thought is that it arises
from the deposition of purely acoustic shock energy generated in the sub-photospheric convection zone \citep{ulmschneider1991}, while 
the absence of appropriate spectroscopic shock signatures in inactive {\it basal-flux} red giants \citep{judge_carpenter1998} suggests a magnetic origin. In fact, on the Sun both mechanisms are seen to operate: small-scale transient shock heating in the 
so-called "K grains" and more  steady magnetic heating in the 
large-scale supergranulation network.

The physical structures of chromospheres from these two mechanisms are likely to be quite different. 
The acoustic wave picture would lead to a 3-D network of shocks that
is highly time variable and at a given position the gas temperature would fluctuate from 
very cool to very hot with the mean temperature being cool \citep{wedemeyerbohm_etal2007}. This 
mechanism leads to localized intermittent chromospheric emission but, which when averaged over the stellar disk, would
not be apparent in the mm-radio.
The presence of magnetic fields can lead to longer lived structures and enhanced 
atmospheric heating. Time-independent 1-D semi-empirical models for the Sun that represent 
regions of different activity are well established (\citealt{val1976,val1981,fal1990}).
Similar but spatially unresolved models have been constructed for a few cool stars including inactive red giants. These semi-empirical models are designed to reproduce the temporally and spatially averaged uv and optical emission line profiles and fluxes. A characteristic of these models is that they require a gradual temperature rise from the top of the photosphere to the upper chromosphere, and then a very steep rise through the chromosphere-coronae transition region \citep{kelch_etal1978, harper1992, luttermoser_etal1994, mcmurry1999}.

One might expect that acoustic models, whose mean temperatures do not increase with
height, would fail to predict the same uv fluxes as semi-empirical models with an outward temperature increase, 
but this is not necessarily the case. In the uv, $h\nu/kT \gg 1$ and the source functions and 
emissivity $\propto e^{-h\nu/kT}$ are very sensitive to temperature. In this situation, 
the shock peaks can dominate the temporally and spatially averaged emission of the dynamic atmosphere. 
Thus it is possible for acoustic shock and semi-empirical models to produce the same chromospheric 
fluxes even though the
temperature structures are profoundly different \citep{carlsson_stein1995}.

However, in the mm-radio, the thermal continuum source (Planck) function depends linearly on $T_{\rm e}$, so both solar and stellar  
mm observations can probe differences
between intermittent and long-term structured atmospheres and potentially differentiate
between acoustic and magnetically heating models. A comparison
between uv and radio diagnostics could test the symbolic inequality
\begin{equation}
\langle e^{-h\nu/kT}\rangle  \ne e^{-h\nu/k\langle T\rangle} 
\label{eq:symbolic_inequality}
\end{equation}
where the averages are over space and time. The LHS exemplifies uv emission
diagnostics while the RHS reflects the same source function but with
the mean temperature, $\langle T\rangle$, as inferred for example from mm radio emission. In practice
uv and mm-radio fluxes calculated from different atmospheric models can be compared to observations. 
The hydrogen ionization determines the electron densities which, in part, determine both the mm-radio opacity and 
the uv collisional excitation rates. The remaining difference in the temperature dependence for formation 
leads to the test of equation (\ref{eq:symbolic_inequality}).

Here we analyze published thermal mm (250~GHz) fluxes \citep{altenhoff_etal1994} from single non-dusty and non-pulsating K and M giants (luminosity class III) to investigate their chromospheric signatures
and explore the role of future multi-frequency observations in revealing the nature of chromospheric structuring. Future solar Atacama Large Millimeter/submillimeter Array (ALMA)imaging will be focussed on obtaining sufficient spatial resolution to separate the magnetic and non-magnetic regions \citep{loukitcheva_etal2008}, whereas
studies of inactive {\it basal-flux} red giants may allow us to 
avoid the contamination from magnetic active regions by observing
giants that are covered with inactive chromospheric regions analogous to 
quiet solar supergranule centres.  

A first estimate for the expected mm flux from a red giant is provided by the
hard disk model; which is appropriate when the ionized density scale height, $H^{\rm ion}_\rho$ is much smaller than the stellar radius, $R_\ast$. The observed flux density ($F_\nu$) is given by
\begin{equation}
F_\nu = \left(\phi/2\right)^2 F_\nu^{\ast},
\end{equation}
where $F_\nu^\ast$ is the average flux density at the surface, having angular diameter $\phi$ with respect to the distance observer. Working in terms of brightness temperatures $T_{\rm Br}$ we have
\begin{equation}
F_\nu= {\pi k \nu^2 \phi^2\over{2c^2}}\int\limits_0^1 T_{\rm Br}\left(\mu\right)\mu\>d\mu = {\pi k \nu^2 \phi^2\over{c^2}}\langle T_{\rm Br} \rangle
\label{eq:flux}
\end{equation}
where $\mu$ is the usual cosine of the angle between a ray and the normal to the emitting surface, and $\langle T_{\rm Br}\rangle$ is the weighted mean brightness temperature.
In more convenient units, the flux in mJy\footnote{1\,mJy$ = 10^{-29}{\rm W\>m}^{-2}{\rm Hz}^{-1}=10^{-26}\>{\rm erg\>s}^{-1}{\rm cm}^{-2}{\rm Hz}^{-1}$} is
\begin{equation}
F_\nu\left({\rm mJy}\right) = 13.8 \left({\langle T_{\rm Br}\rangle\over{3900\>{\rm K}}}\right)
\left({\nu\over{250\>{\rm GHz}}}\right)^2
\left({\phi\over{10\>{\rm mas}}}\right)^2
\label{eq:convenient}
\end{equation}
Many nearby red giants now have measured angular diameters, $\phi_\ast$, and estimated
effective temperatures, $T_{\rm eff}$ \citep{mozurkewich_etal2003}. Inserting $\phi_\ast$ and 
substituting $T_{\rm eff}$ for $\langle T_{\rm Br}\rangle$ into equation (\ref{eq:convenient}) shows that many of these giants  can in principle be observed at multiple frequencies with high signal-to-noise ratios when ALMA is completed given the expected sensitivity limits \citep{butler_wooten1999}. Here we develop a more physically motivated semi-analytical model for  mm fluxes that incorporates constraints on chromospheric structure from independent IR and uv 
observations, which also allows us to include the effects of different levels of chromospheric heating.

In \S2 we compile pertinent empirical constraints derived from semi-empirical
models and other studies of red giants; and in \S3  estimate the expected chromospheric mm optical 
depths. In \S4 we develop
a semi-analytic model for $\langle T_{\rm Br} \rangle$ based on these empirical constraints,
which we then calibrate against direct measurements of the well studied red giant $\alpha$~Tau (K5~III) (Aldebaran).
A comparison with observations and specific predictions for ALMA frequencies 
are presented in \S5, and a discussion of the diagnostic potential of  mm observations 
in \S6. Conclusions are drawn in \S7.

\section{Empirical Constraints on Stellar Chromospheres}

In this section we gather together nuggets of information from previous studies of red giant
chromospheres, which we will incorporate into our analytic model for the  mm radio fluxes.
The radio optical depth, $\tau_\nu$, is proportional to different powers of the 
atmospheric properties - electron density, temperature,
and path length - and we seek to estimate these for stars covering a range of $T_{\rm eff}$ and different levels of
stellar activity.

Historically, the usual lack of temporal and spatial resolution in observations of the chromospheres of 
distant stars partly motivated the development of 1-D time-independent semi-empirical models.
The earliest models used diagnostics observable from the ground -- 
Ca~II H \& K, the Ca~II IR triplet, and H$\alpha$ -- and early satellite observations of Mg~II h \& k
from {\it Copernicus} \citep{ayres_etal1974,ayres_linsky1975}. These types of models are analogous to
the venerable VAL series of solar models, although for the latter spatial resolution did permit
construction of components representative of regions of different activity levels
\citep{val1976,val1981}. More recent variants of these solar component models are the FAL series
\citep{fal1990}, which are employed in solar irradiance studies \citep{fontenla2009}, among other uses.
Two notable features of these semi-empirical models are: (1) the temperature inversion in the
low chromopshere above a temperature minimum, 
$T^{\rm uv}_{\rm min}\simeq 0.75 T_{\rm eff}$; and (2) that the electron density, $n_{\rm e}$, is, to within a factor of $\sim 2$, approximately constant throughout the chromosphere \citep{ayres1979,harper1992}. At the temperature minimum, the electron density ($\sim 10^{-4}n_{\rm H}$) is dominated by (photo)ionized metals; but as the temperature increases through the chromosphere, hydrogen gradually becomes more ionized until at the top of the chromosphere ($\sim 10^4$\,K), hydrogen is predominantly ionized and then $n_{\rm e}\sim n_{\rm H}$.

Thus, through the chromosphere, the increasing ionization of hydrogen counteracts the rapid outward decline of the density due to hydrostatic equilibrium, maintaining a more-or-less constant $n_e$ with increasing altitude \citep{ayres1979}. However, once the hydrogen has become mostly ionized, there no longer is a ready source of bound electrons to be liberated, and so in the higher layers the electron density will fall off rapidly, following the hydrostatic decline in $n_{\rm H}$ itself \citep{ayres1979}.

 While there is a debate
on whether the Sun has a permanent chromospheric temperature rise or 
not - it should be noted that
presently the semi-empirical models nevertheless are the {\em best 
thermodynamic inventory of stellar chromospheric plasma} available.
They approximately reproduce the amounts of plasma at different electron temperatures. This is discussed further in \S6.

Pre-{\it IUE}, Mg~II h \& k observations revealed a strong dependence of chromospheric heating with effective temperature, $F_{\rm Mg II}^\ast \sim T_{\rm eff}^6$ \citep{linsky_ayres1978,ayres1979}. 
\citet{ayres1979} developed a scaling law for the mass column density at the onset of the 
chromospheric temperature rise, which can be rephrased in terms of the
chromospheric electron density, $n_{\rm e}$, where atmospheric heating is sufficient to overcome $H^-$ cooling, namely 
\begin{equation}
n_e \sim {\cal A}_{\rm Fe}^{1/2} {\cal F}^{1/2} g_\ast^{1/2} T_{\rm eff}^{5/2}.
\label{eq:ayres}
\end{equation}
${\cal A}_{\rm Fe}$ represents the abundance of low first ionization potential elements relative to solar, ${\cal F}$ is
the possible enhancement of a particular star relative to the general $T_{\rm eff}$ scaling related to differences in intrinsic activity. In \citet{ayres1979}, ${\cal F}=1$ was taken
to represent levels appropriate to inactive stars or the quiet sun, and ${\cal F}=10$ for solar plage regions or (fast rotating) stars showing enhanced activity. Some empirical evidence for this trend on $n_{\rm e}$ in red giants is given by \citet{byrne_etal1988} based on {\it IUE} C~II] 2325\,\AA {} intersystem flux ratios, and by \citet{buchholz_etal1998} who compared time-dependent acoustically heated atmospheric models with the mass column density of an effective $T^{\rm uv}_{\rm min}$. Chromospheric mm radio fluxes are sensitive to the opacity, which 
in turn depends on the square of the electron density. Here we assume the scaling in equation~(\ref{eq:ayres}) provides
an appropriate dependence of $n_{\rm e}$ on star specific parameters.   

\subsection{Basal Fluxes}
We begin by considering the ratio of a star's measured Mg~II h \& k flux to the {\it basal-flux}, i.e., a {\it basal-flux} star has ${\cal F}=1$.
We recall the usual implicit assumption that Mg~II h \& k line fluxes represent a fixed
fraction of the total chromospheric radiative losses \citep{linsky_ayres1978}. The other major coolants are H~Ly$\alpha$, Ca~II, and
Fe~II (e.g., \citealt{judge_stencel1991}).

For this work we take {\it IUE} Mg~II h \& k fluxes from \citet{martinez_etal2011}, which have typical  uncertainties of $\pm 10$\%, and we also adopt their $V-K$-based expression for the 
{\it basal-flux} (erg~cm$^2\>{\rm s}^{-1}$),  
\begin{equation}
\log{F_{\rm basal}} = 6.78\log{T_{\rm eff}} - 19.74
\label{eq:martinez}
\end{equation}
which has a very similar $T_{\rm eff}$-dependence to \citet{linsky_ayres1978}.
We have chosen specifically their $V-K$ form of the {\it basal-flux}, because the implied angular diameters are 
more reliable than the $B-V$ form at lower $T_{\rm eff}$. \cite{judge_stencel1991} presented Mg~II h \& k surfaces fluxes
for a sample of low and intermediate mass giants, and their lower-bound drops below equation (\ref{eq:martinez}) 
for $T_{\rm eff} < 3300$~K.

\subsection{Inhomogeneous Chromospheres}

Observations of the $4.6\mu$m CO fundamental bands have demonstrated that on the Sun and late-type stars
there exist regions of cool molecular gas with temperatures well below the $T^{\rm uv}_{\rm min}$ derived from ultraviolet and optical diagnostics \citep{ayres_testerman1981, wiedemann_etal1994}. On the sun the area filling factor of the CO material has been proposed to be up to 80\% in the low chromosphere \citep{ayres2002}, but that factor is controversial.
For the red giants the filling factor is close
to unity with the temperature scaling as $T_{\rm CO} \sim 0.60 T_{\rm eff}$ (which is close to radiative equilibrium values when molecular cooling is included). At present there is a vigorous debate as to the exact nature of the CO clouds
given the highly dynamic, time-dependent nature of the solar atmosphere
 \citep*{kalkofen_etal1999,ayres2002,leenaarts_etal2011}.
The presence of the CO material must be accounted for in chromospheric models, particularly when these regions are probed by mm continuum emission. For example, recent spectral interferometry of 
the much later spectral-type M giant BK~Vir (M7 III) has revealed the presence of spatially extended CO with an inner extended component at $R=1.2R_\ast$, with $T_{\rm CO}\simeq 0.65 T_{\rm eff}$ \citep{ohnaka_etal2012}.

At lower effective temperatures, other molecules, such as water vapour \citep{tsuji2008}, can play a role in further reducing the effective boundary layer temperature. \citet{quirrenbach_etal1993} found evidence from the Mark III
interferometer that the TiO continuum becomes significantly geometrically extended by M5~III (R Lyr),
for which the ratio of TiO to continuum diameters is $1.12\pm 0.02$. These extended molecular components possibly are related to the `molsphere phenomenon' that is more common in the Mira and lower gravity M stars
\citep{tsuji_etal1997,perrin_etal2004}.

\subsection{Chromospheric Extent}

The apparent angular size of a stellar radio source clearly is an important determinant of the radio flux. Here we consider
the factors that control the angular diameter in the  mm band, and in particular its ratio,
$G$, to the photospheric angular diameter, i.e., $\phi_{\rm mm}=G\phi_\ast$.

The ratio of pressure scale-height, $H_{\rm pres}$, to 
stellar radius, $R_\ast$, is 
\begin{equation}
{H_{\rm pres}\over{R_\ast}} \sim {T_{\rm eff} R_\ast\over{M_\ast}}, 
\end{equation}
so that red giants with their much larger radii, but near solar masses, 
have relatively thicker chromospheres than the Sun.

Eclipse observations of the $\zeta$~Aurigae systems, which have 
intermediate mass K~Ib primaries, show large plasma turbulence and 
direct evidence for atmospheric extensions greater than the thermal 
pressure scale height, (e.g., \citealt{eaton1993}). This indicates
that $H_{\rm pres}$ is controlled by both thermal and non-thermal 
motions. Here we describe the increase in $H_{\rm pres}$
from non-thermal flows by a quasi-hydrostatic equilibrium, 
\begin{equation}
H_{\rm pres} = {v^2 + 2kT\over{2 \langle\mu\rangle m_{\rm H} k g_\ast}},
\label{eq:presturb}
\end{equation}
where $v$ is a characteristic velocity. For $\zeta$~Aur systems, $v$ is
consistent with the turbulence inferred from spectra, and possibly
correlated on spatial scales comparable to the density/pressure scale-height. 
$\langle\mu\rangle$ is the chromospheric average of the mean mass per particle in units of the hydrogen mass $m_{\rm H}$, and $g_\ast$ is the 
surface gravity. We adopt $\langle\mu\rangle=1.2$ corresponding to a partially ionized hydrogen ($n_{\rm e} \simeq 0.1n_{\rm H}$) and surface helium abundance of $n_{\rm He}=0.1 n_{\rm H}$. 
Semi-empirical models typically include this non-thermal term in the equations of
quasi-hydrostatic equilibrium. For $\zeta$~Aur systems, the implied turbulence is slightly greater than the hydrogen sound speed, and in our model we set $v=\sqrt{2kT}$. 
While estimates of stellar radii are available from interferometry (e.g., \citealt{baines_etal2010}) and
nearby red giants also have {\it Hipparcos} parallaxes, the stellar masses 
remain uncertain for the single stars. Typical estimates are $1-2M_{\odot}$, leading to some uncertainty in the surface gravity in equation~(\ref{eq:presturb}).

The semi-empirical one-component model of $\alpha$~Tau \citep{mcmurry1999} predicts a 250~GHz (1.2\,mm) radial optical depth of unity at $R\simeq 1.03R_\ast$, but the 
apparent angular diameter (half central intensity) corresponds to $\phi_{\rm mm} \simeq 1.1\phi_\ast$. This is a result
of the larger column density of the tangential sight-lines 
that define the 1.2\,mm limb.
The apparent angular diameter of $\alpha$~Boo predicted from a semi-empirical model of \citep{drake1985} is 
$\phi_{\rm mm} \simeq 1.06\phi_\ast$, and for g~Her (M6~III) $\phi_{\rm mm} \simeq 1.08\phi_\ast$ \citep{luttermoser_etal1994}. These models
were constructed with different diagnostics and assumptions
about the contribution of turbulent pressure. So although there is a trend of increasing fractional 
chromospheric extension with later spectral type, the lower $T_{\rm eff}$ and thus lower chromospheric heating,
leads to lower $n_{\rm e}$ and reduced mm-optical thickness. In light of these
competing factors we adopt a typical extension of 
$1.08\phi_\ast$ (i.e., $G=1.08$), and accept an additional uncertainty of $\sim 2\%$ in $G$ which is consistent with the level of approximations adopted here. In summary, here $\langle T_{\rm Br}\rangle$ for the  mm radio is defined as
\begin{equation}
\langle T_{\rm Br}\rangle = {2 c^2 F_\nu\over{\pi k \nu^2 \left(G\phi_\ast\right)^2}}
\label{eq:}
\end{equation}

Very recently, single baseline, spectrally-resolved, K giant chromospheric 
Ca~II IR-triplet visibilities from CHARA have been presented by \citet{berio_etal2011}. 
The authors attempted to estimate the chromospheric extent by using a correction procedure based on the \citet{eriksson_etal1983} semi-empirical chromosphere of the coronal star $\beta$~Cet (K0~III). \cite{berio_etal2011} derive chromospheric extents of $1.16-1.47R_\ast$, although these values are based on a model whose actual spatial extent is only $1.02R_\ast$. The extended emission observed in the cores of the IR-triplet lines depends on the presence of Ca~II ions which might not be related to the regions of higher electron density to which the radio is sensitive. The calculation of the IR-triplet visibilities requires a careful treatment of the Ca~II to Ca~III photoionization by H~Ly$\alpha$
\citep{linsky_avrett1970,rowe1992}, spherical geometry, and possible impact of cross-redistribution
of photons scattering within the Ca~II H \& K resonance lines \citep{uitenbroek1989}. The new CHARA measurements are very important diagnostics and one of us (NO'R) is undertaking these non-LTE radiative transfer calculations at the present time.

\begin{figure*}
  \includegraphics[scale=0.5,angle=90]{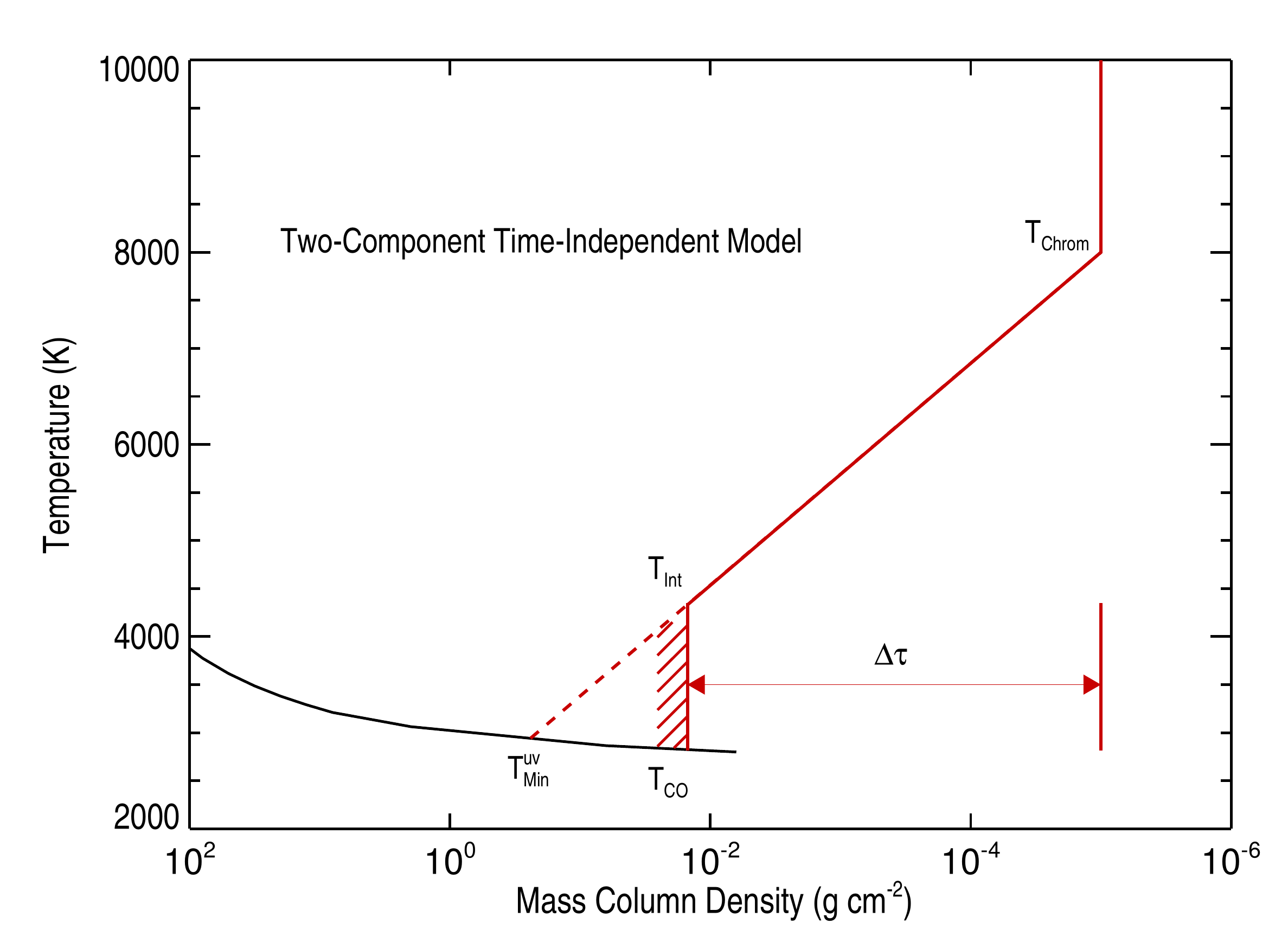}
  \vspace*{0cm}
  \caption{Schematic diagram of an idealized static two-component inhomogeneous red giant chromosphere.
  The warm chromosphere temperature rises outwards from $T_{\rm Int}$ to $T_{\rm Chrom}$ and had a total optical thickness $\Delta\tau$. Beneath the warm component, lies an opaque cool molecular region with $T_{\rm CO} < T^{\rm uv}_{\rm Min}$.}
\end{figure*}

\begin{figure*}
  \includegraphics[scale=0.5,angle=90]{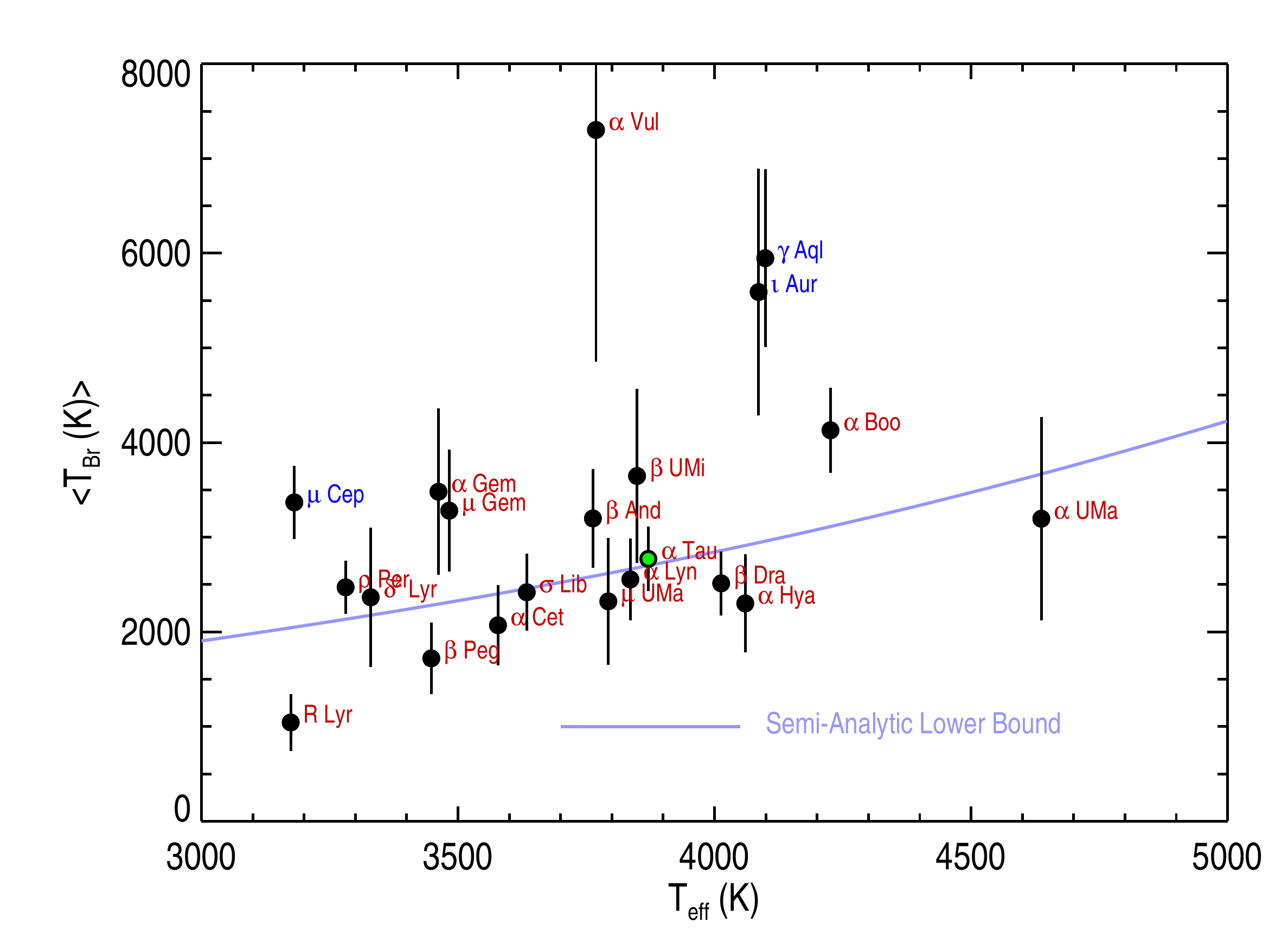}
  \vspace*{0cm}
  \caption{250~GHz brightness temperatures for the sample given in Table~1. We also show three additional stars ($\mu$~Cep, $\iota$~Aur, and $\gamma$~Aql) which are of higher luminosity and might have emission in excess to that predicted by our model. The basal $\langle T_{\rm Br}\rangle$ predicted from our model is also
  shown as a solid curve.}
\end{figure*}

\section{Chromospheric millimetre optical thickness}

As mentioned earlier the chromospheric electron density has contributions from the
low first ionization potential elements and from hydrogen. At low temperatures ($T < 6000~K$), 
ionization of hydrogen is a two stage process: excitation to the $n=2$ level by electron collisions or scattered H~Ly$\alpha$ photons followed by photoionization by the optically thin photospheric Balmer
continuum. At higher temperatures, direct collisional ionization becomes important. When hydrogen is partially ionized, free-free (thermal bremsstrahlung) opacity 
dominates at  mm wavelengths. To examine the influence of the stellar outer atmosphere 
on the mm
thermal emission, we first consider the optical thickness of
the warm chromospheric component. For typical conditions, the opacity corrected for
stimulated emission is given by \citet{rybicki_lightman2004}
\begin{equation}
\kappa = {0.01772\> n_{\rm e} n_{\rm ion}Z_{\rm ion}^2 \over{T_{\rm e}^{1.5}\nu^{2}}} g_{\rm ff}  \>\>{\rm cm}^{-1}
\end{equation}
where the variables are in cgs units, $Z_{\rm ion}$ is the charge of the ions, $n_{\rm e}$ and $n_{\rm ion}$ are the number densities of the electrons and ions, respectively, and $g_{\rm ff}$ is the free-free Gaunt factor. At cm-wavelengths the familiar power-law
approximation from \citet{altenhoff_etal1960} is
\begin{equation}
g^{\rm cm}_{\rm ff} = 11.96 T_{\rm e}^{+0.15}\nu^{-0.1}, 
\end{equation}
but at ALMA frequencies,which range from 84-950 GHz (Band 3-10), the
power-law dependence is slightly different, namely
\begin{equation}
g^{\rm sub-mm}_{\rm ff} = 24.10 T_{\rm e}^{+0.26}\nu^{-0.17},
\end{equation}
as derived using the accurate Gaunt factors from \citet{hummer1988}.
Under typical chromospheric conditions, the majority of abundant species are either neutral or single ionized so that $Z_{\rm ion}=1$ and $n_{\rm e}=n_{\rm ion}$. This leads to
an expression for the  mm opacity,
\begin{equation}
\kappa_\nu = {0.427 n_{\rm e}^2 \over{T_{\rm e}^{1.24}\nu^{2.17}}}  \>\>{\rm cm}^{-1}
\end{equation}
The optical thickness of a slab of depth, $\Delta L$, is then simply $\Delta\tau_\nu=\kappa_\nu \Delta L$.

The assumption that the electron density is approximately constant, as the
ionization $n_{\rm e}/n_{\rm H}$ goes from $10^{-4}\to 1$, implies that the hydrogen density
has declined by four orders of magnitude and therefore
the physical thickness of the layer $\Delta L \simeq 9$ density scale-heights. To the
level of approximation here we set $H_{\rho}\simeq H_{\rm pres}$

For the radial
optical depth of the chromosphere we have
\begin{equation}
\Delta\tau_\nu \simeq {0.427\>n_{\rm e}^2\over{T_{\rm e}^{1.24}\nu^{2.17}}}\cdot 9 \cdot
{2 k T_{\rm e}\over{\langle \mu\rangle m_{\rm H} g_\ast}} \simeq {6.4\> n_{\rm e}^2\over{T_{\rm e}^{0.24}\nu^{2.17}}}
{k \over{m_{\rm H} g_\ast}}
\label{eq:dtau}
\end{equation}

This expression reveals only a very weak dependence on temperature variations within
the chromosphere which is a result of the cancellation resulting from the 
combination of opacity and scale-height. In the following, we set the surviving
temperature factor to $T_{\rm eff}^{0.24}$. Inserting into equation~(\ref{eq:dtau}) typical values for the 
M2.5~III star $\beta$~Peg (\citealt{decin_etal2003}: $M_\ast=2.0M_\odot$, $R=110R_\ast$, $T_{\rm eff}=3600$~K)
and a typical uv-based value of $n_{\rm e}\sim 6\times 10^8\>{\rm cm}^{-3}$ \citep{byrne_etal1988}, we find
for 1.2\,mm (250~GHz) and 3\,mm (100~GHz) chromospheric optical thicknesses of $\Delta\tau_{{\rm 250 GHz}}\simeq 0.1$  and $\Delta\tau_{{\rm 100 GHz}}\simeq 0.8$, respectively. Thus depending
on the level of stellar activity, the  mm wavelength regime is one where chromospheres
can go from optically thick at mm and cm wavelengths to optically thin in the sub-mm.

If the optical thickness of the chromosphere is less than unity the next emitting layer we must consider
is the cool molecular gas immediately above photosphere with its near unity filling factor. In these cool, high density layers, where the hydrogen density increases rapidly with depth, H$^-$ free-free can become an important opacity source \citep{harper_etal2001}. Here we 
assume that the bifurcated molecular regions are opaque at these wavelengths. Thus the optical depth unity is reached when $T_{\rm e} \simeq T_{\rm CO}$, which is reasonable if the temperatures in these outer radiative equilibrium layers change only slowly with optical depth.

\begin{table*}
\begin{minipage}{126mm}
\caption{Adopted properties of $\alpha$~Tau.}
\begin{tabular}{llll}
\hline  
 Quantity         & Value & Unit  & Reference \\
 Spectral Type    &       & K5~III &           \\ 
 $\phi_\ast$      & $21.1\pm 0.2$  & mas & \citet{mozurkewich_etal2003}  \\
 $T_{\rm eff}$        & $3871\pm 48$   & K & ditto  \\ 
 $R_\star$        & $44.56\pm 2.94$& $R_\odot$ & \citet{decin_etal2003} \\
 $\log{g_\ast}$   & $1.5\pm 0.15$ & ${\rm cm\>s}^{-2}$ & ditto\\
 $M_\star$        & $2.30\pm 0.85$ & $M_\odot$ & inferred ($g_\star$) \\
 {[Fe/H]}         & $-0.15\pm0.20$ &  & ditto  \\
 $F_{\rm Mg~II}(h+k)$ & $6.50\times 10^4$ & ${\rm erg\>cm}^{-2}{\rm s}^{-1}$ & \citet{martinez_etal2011}+$\phi_\ast$ \\ 
 $F_{\rm Mg~II}/F_{\rm basal}$ & 1.69 & & ditto \\ 
 \hline
\end{tabular}
\end{minipage}
\end{table*}

\section{Description of the Semi-Analytic Model}

We now assemble the different aspects discussed above and construct a semi-analytic
model to provide estimates of the  mm, e.g., ALMA band, radio fluxes. We then calibrate against well studied $\alpha$~Tau (K5~III) to normalize all the approximation constants of order unity accumulated along the way. A schematic representation of the model
inhomogeneous chromosphere is shown in Figure 1. Above the uv temperature minimum $T_{\rm Min}^{\rm uv}$ lies the opaque
cool molecular material with temperature $T_{\rm CO}$. Immediately above this, perhaps in an overlying magnetic canopy, is the base of the warm chromosphere with temperature $T_{\rm Int}$; $T_e$ then rises
linearly in mass column density to the upper chromosphere with temperature $T_{\rm Chrom}$ and has an
optical thickness $\Delta\tau$. 
In the model $n_{\rm e}$ is approximately constant (to within a factor of $\pm 2$) 
and $T_{\rm e}$, in the warm component, is approximately linear with height. The source function $S_\nu \propto T_{\rm e}$, and 
since the  mm optical depth is $\tau_\nu \propto n_{\rm e}^2 \Delta L$, then $S_\nu$
is a linear function of $\tau_\nu$. In this case the integration of the formal solution of the radiative transfer equation for the brightness temperature can be done
analytically.

$T_{\rm Br}$, a function of frequency and viewing angle, represents a weighted contribution of the three temperature parameters and is given by
\begin{equation}
T_{\rm Br}\left(\mu\right) = \alpha T_{\rm CO} + \gamma T_{\rm Int} + \beta T_{\rm Chrom}
\label{eq:tbr}
\end{equation}
 where $T_{\rm CO}, T_{\rm Int}, T_{\rm Chrom}$ are the key temperatures described above and weighting coefficients are
 obtained from the formal solution of the transfer equation \citep{olson_kunasz1987}, which in our notation are
\begin{equation}
\alpha = e^{-\Delta\tau}
\label{eq:a}
\end{equation}

\begin{equation}
\beta =  1 - \left(1 - \alpha\right)/\Delta\tau
\label{eq:b}
\end{equation}

\begin{equation}
\gamma = 1 - \alpha - \beta.
\label{eq:c}
\end{equation}

From \citet{wiedemann_etal1994} we adopt $T_{\rm CO}\simeq 0.6T_{\rm eff}$ and $T_{\rm Int}=4500$\,K. We take $T_{\rm Chrom}=10000$\,K where hydrogen has become 50\% ionized \citep{mcmurry1999} and any further increase in temperature now leads to a rapid decline in $n_{\rm e}$ and hence also $\kappa_\nu$, creating a transparent boundary above $T_{\rm Chrom}$.

In this analytic model, the radio fluxes are dependent on $T_{\rm eff}$ and the
optical depth of the warm chromosphere layer above the
CO zone. 
If the emitting layers are not too thick compared to the stellar radius,
the stellar flux can be described as the integral of $T_{\rm Br}\left(\mu\right)$ for a plane-parallel slab, where $\mu=\cos\theta$ and $\theta$ is the angle between the atmosphere normal and the ray.  This allows us to write a weighted mean brightness temperature,
e.g., in equation (\ref{eq:convenient}), as
\begin{eqnarray}
\langle T_{\rm Br}\rangle & = & 2 \int\limits_0^1 T_{\rm Br}\left(\mu\right) \mu\> d\mu \cr
& = &
2 \sum\limits_{i=1}^{n} T_{\rm Br}\left(\mu_i\right) \mu_i w_i = T_{\rm Br}\left(\mu=1/2\right)
\end{eqnarray}
In keeping with the
simple nature of our analytic model we replaced the integral by a single point Gaussian quadrature 
($n=1$, $\mu_1=1/2$, and $w_1=1$). $\langle T_{\rm Br}\rangle$ is found by substituting $\Delta\tau$ from equation~(\ref{eq:dtau}), adjusted for the viewing angle $\mu=1/2$, into equations~(\ref{eq:a}), (\ref{eq:b}), and (\ref{eq:c}) to evaluate the
weighting coefficients and then solve for equation~(\ref{eq:tbr}). 

\subsection{Calibrating with $\alpha$~Tau (K5~III)}

To mitigate some of the inherent uncertainties in the model we 
evaluate the chromospheric optical thickness (equation~\ref{eq:dtau}) for 
$\alpha$~Tau using empirical constraints. 
We initially use the scheme outlined above to 
calculate the 250~GHz flux and compare it to the measured 1986 
IRAM \cite{altenhoff_etal1994} flux of
$51\pm 6$\,mJy. $\Delta\tau$ then is scaled by a constant to normalize the model. We use the \cite{altenhoff_etal1994} compilation because it represents the largest stellar sample available. 
We later
(\S5.3) compare $\alpha$~Tau
to more recent interferometric observations by \cite{cohen_etal2005}. 

\citet*{robinson_etal1998} give estimates of $n_{\rm e}$ in $\alpha$~Tau for three epochs (1990, 1994, and 1996)
based on near-uv C~II spectra obtained with the {\it Hubble Space Telescope}: $n_{\rm e} \simeq 10^9\>{\rm cm}^{-3}$.  Unfortunately there were no contemporaneous
{\it IUE} observations in 1986 when the IRAM 250~GHz observations were made. 
Electron densities inferred from the C~II] 2325\,\AA {}
multiplet ratios are weighted towards regions of high $T_{\rm e}$ and $n_{\rm e}$. Because
semi-empirical models show that $n_{\rm e}$ is constant to within a factor of $\sim 2$,
we adopt a mean $n_{\rm e}$ that is half the peak observed value, $<n^2_{\rm e}>=n^2_{\rm e}(HST)/4$.
 
The predicted flux is then obtained from equation~(\ref{eq:flux}) with $G=1.08$ and $\langle T_{\rm Br}\rangle$
given by the model. Combining the stellar parameters adopted in Table~1, we derive 
$F_{\rm 250GHz}=52$\,mJy, likely a fortuitous agreement. The model predicts a radial optical depth of $\Delta\tau_{\mu=1}=0.05$ and $\Delta\tau_{\mu=1/2}=0.09$. 

\subsection{Star to star variations in $\Delta\tau$}

$\alpha$~Tau's Mg~II h \& k flux is about 1.7 times basal, and it might be slightly metal poor (see Tables 2 and 1, respectively) both of which affect the optical thickness.
To scale this chromospheric thickness to other red giants, and
remove explicit dependence on $\alpha$~Tau's stellar parameters, we define an
optical thickness $\Delta\tau_\nu\left(\mu\right)$ for a fictitious star with $\alpha$~Tau's
$T_{\rm eff}$ but now with basal-flux heating (${\cal F}=1$) and solar metalicity.
Using the \citet{ayres1979} scaling of equation~(\ref{eq:ayres}) gives $\Delta\tau_{\mu=1}^{\rm basal}=0.04$.

Combining equation~(\ref{eq:ayres}) and equation~(\ref{eq:martinez}), the chromospheric optical thickness 
becomes
\begin{equation}
\Delta\tau_\nu^\ast\left(\mu\right) = {0.04\over{\mu}}
{A_{\rm Fe}\over{A_{\rm Fe\odot}}}{F_\ast\over{F_{\rm basal}}}
\left({T_{\rm eff}\over{3871\>{\rm K}}}\right)^{5.54}
\left({\nu\over{250\>{\rm GHz}}}\right)^{-2.17}.
\label{eq:dtaustar}
\end{equation}
Note that the surface gravity term has now cancelled.
\begin{table*}
\begin{minipage}{126mm}
\caption{Stellar properties, 250 GHz radio fluxes, and chromospheric optical thicknesses. 
The stellar properties are from Mozurkewich et al. (2003),
the mm fluxes, except $\sigma$~Lib, are from Altenhoff et al.(1994) and 
the Mg~II fluxes are from Martinez et al. (2011).}

\begin{tabular}{rrrlccrrrrcc}
\hline
 Star    & HR   &   HD   & Spectral & $T_{\rm eff}$ & $\sigma(T_{\rm eff})$ & $\phi$ & $\sigma(\phi)$ & $Flux$ & $\sigma(Flux)$ & $F/F_{\rm basal}$ & $\Delta\tau$ \\
 & & & Type & $(K)$ & $(K)$ & $(mas)$ & $(mas)$ & (mJy) & (mJy) & (Mg~II h \& k) & ($\mu=1/2$)  \\
\hline
$\beta$~And  &  337 &   6860 &     M0~III & 3763 &  46 & 13.75 & 0.14 & 25.0 & 4.0 &  2.2 &  0.15 \\
$\alpha$~Cet &  911 &  18884 &  M1.5~IIIa & 3578 &  53 & 13.24 & 0.26 & 15.0 & 3.0 &  2.0 &  0.10 \\
$\rho$~Per   &  921 &  19058 &      M4~II & 3281 &  40 & 16.55 & 0.17 & 28.0 & 3.0 &  1.7 &  0.05 \\
$\alpha$~Tau & 1457 &  29139 &     K5~III & 3871 &  48 & 21.10 & 0.21 & 51.0 & 6.0 &  1.7 &  0.09 \\
$\iota$~Aur  & 1577 &  31398 &      K3~II & 4086 &  50 &  7.50 & 0.07 & 13.0 & 3.0 &  1.7 &  0.18 \\
$\eta$~Gem   & 2216 &  42995 &     M3~III & 3462 &  43 & 11.79 & 0.12 & 20.0 & 5.0 &  3.0 &  0.13 \\
$\mu$~Gem    & 2286 &  44478 &   M3~IIIab & 3483 &  43 & 15.12 & 0.15 & 31.0 & 6.0 &  1.3 &  0.06 \\
$\alpha$~Lyn & 3705 &  80493 &     K7~III & 3836 &  47 &  7.54 & 0.07 &  6.0 & 1.0 &  1.9 &  0.14 \\
$\alpha$~Hya & 3748 &  81797 &  K3~II-III & 4060 &  50 &  9.73 & 0.10 &  9.0 & 2.0 &  1.3 &  0.14 \\
$\mu$~UMa    & 4069 &  89758 &     M0~III & 3793 &  47 &  8.54 & 0.09 &  7.0 & 2.0 &  2.4 &  0.17 \\
$\alpha$~UMa & 4301 &  95689 &     K0~III & 4637 &  62 &  6.74 & 0.10 &  6.0 & 2.0 &  1.0 &  0.22 \\
$\alpha$~Boo & 5340 & 124897 &     K2~III & 4226 &  53 & 21.37 & 0.25 & 78.0 & 8.0 &  2.7 &  0.11 \\
$\beta$~UMi  & 5563 & 131873 &     K4~III & 3849 &  47 & 10.30 & 0.10 & 16.0 & 4.0 &  2.2 &  0.12 \\
$\gamma$~Dra & 6705 & 164058 &     K5~III & 4013 &  52 &  9.86 & 0.13 & 10.1 & 1.3 &  1.2 &  0.12 \\
$\delta^2$~Lyr & 7139 & 175588 &   M4~II  & 3330 &  44 & 11.53 & 0.16 & 13.0 & 4.0 &  1.9 &  0.07 \\
   R~Lyr     & 7157 & 175865 &     M5~III & 3174 &  41 & 18.02 & 0.22 & 14.0 & 4.0 &  1.5 &  0.04 \\
$\gamma$~Aql & 7525 & 186791 &     K3~II  & 4099 &  50 &  7.27 & 0.07 & 13.0 & 2.0 &  2.2 &  0.24 \\
$\beta$~Peg  & 8775 & 217906 &   M2.5~III & 3448 &  42 & 17.98 & 0.18 & 23.0 & 5.0 &  1.6 &  0.07 \\
$\sigma$~Lib & 5603 & 133216 &  M3.5~III   & 3634 & 110 & 11.00 & 0.05 & 12.1 & 2.0 &  2.0 &  0.11 \\
\hline
\end{tabular}
\end{minipage}
\end{table*}

\begin{figure*}
  \includegraphics[scale=0.5,angle=90]{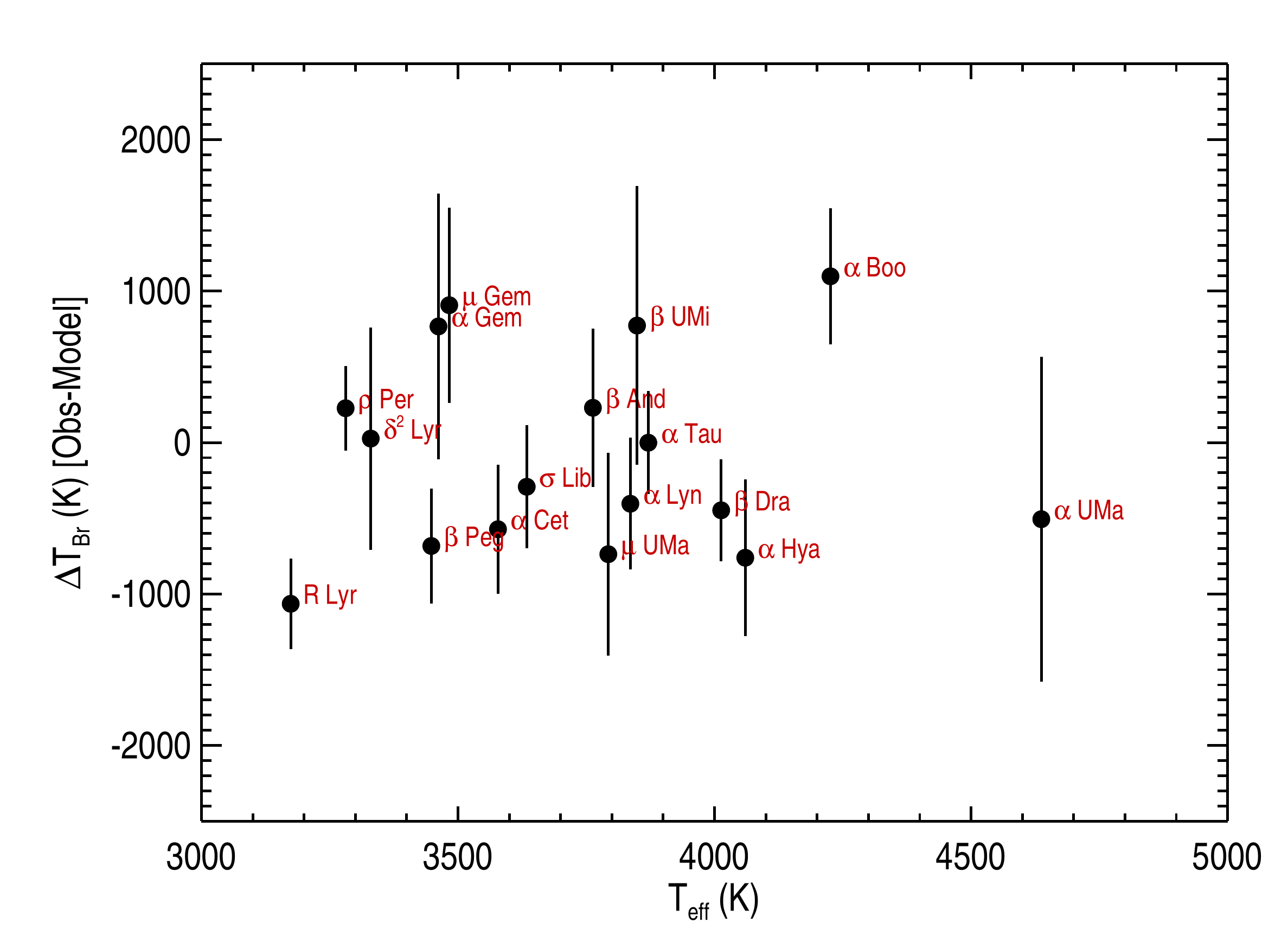}
  \vspace*{0cm}
  \caption{Observed minus model 250~GHz brightness temperatures. The model has been calibrated    against $\alpha$~Tau and the uncertainties
  in the Mg~II fluxes have not been included in the errors shown. While the model
  is consistent with the existing data, new higher accuracy observations are required to
  assess the model's viability as a  mm flux estimator.}
\end{figure*}

\begin{figure*}
  \includegraphics[scale=0.5,angle=90]{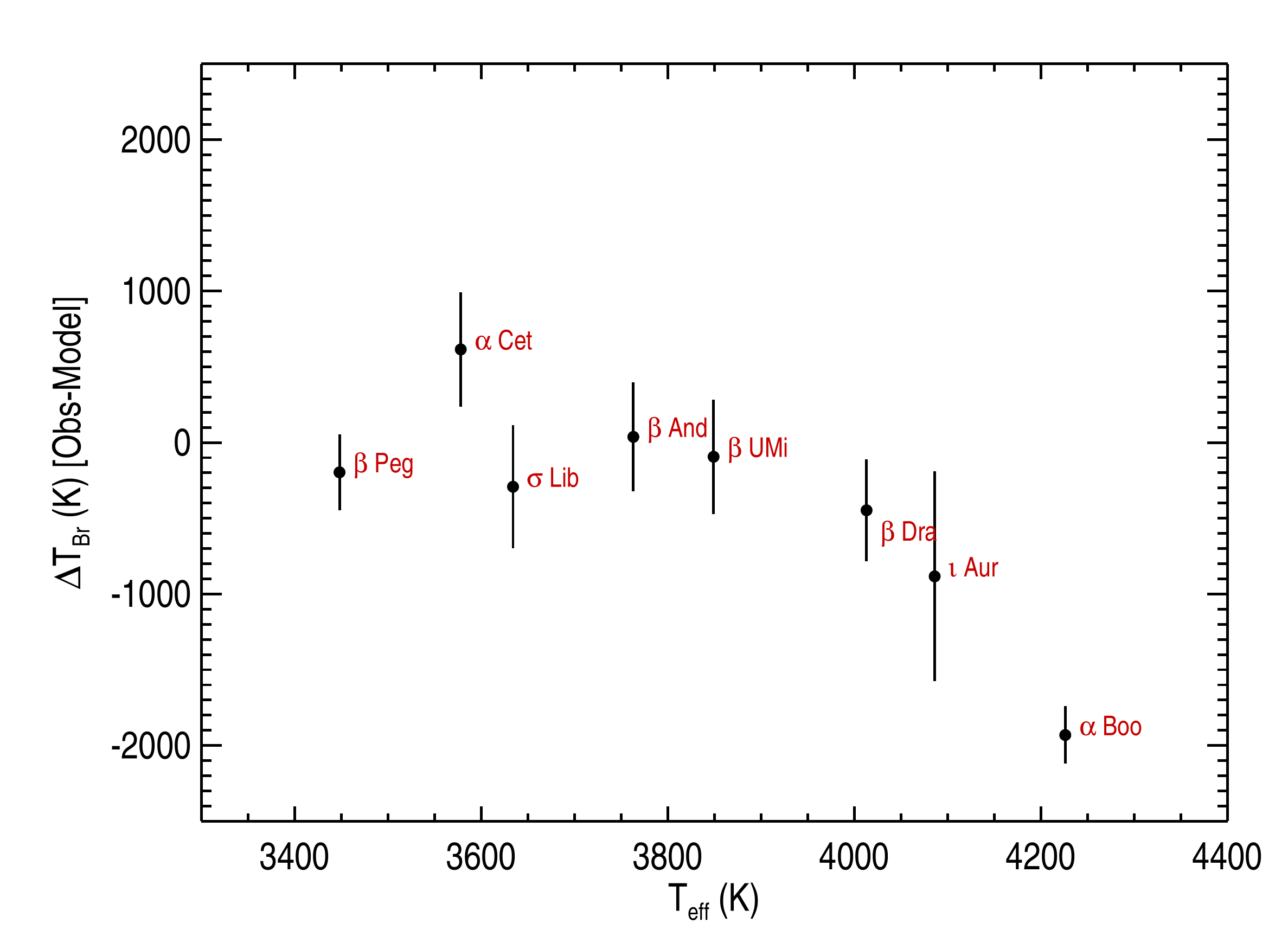}
  \vspace*{0cm}
  \caption{Dehaes et al. (2011) 1.2\,mm MAMBO II: Observations minus model brightness temperatures plotted
  with same y-axis as used in Figure 3. The SIMBA fluxes for $\alpha$~Boo and $\alpha$~Cet lie outside this range. The model is consistent with these higher S/N fluxes, but note Dehaes et al. discount the MAMBO II fluxes for $\alpha$~Boo and $\alpha$~Cet owing to unsuitable weather conditions.}
\end{figure*}

\section{Comparison with Observations and Predictions}

To test our semi-analytic model for the  mm thermal fluxes, we require
a sample of single inactive {\it mundane} red giants with well determined stellar parameters. 
The 250~GHz (1.2\,mm) single 30\,m dish IRAM survey of \citet{altenhoff_etal1994} presents the largest
sample of measurements for these stars. We combined this survey with the accurate $\phi_\ast$ and $T_{\rm eff}$ sample from the Mark III optical interferometer \citep{mozurkewich_etal2003}. The stars in common 
are listed in Table~2. We also have added $\sigma$~Lib (M3~III) from the smaller sample of \cite{dehaes_etal2011}. This combination of data enables us to make initial tests
of the model. In Figure 2, we compare the observed $\langle T_{\rm Br}\rangle$ (G=1.08), with the predicted {\it basal-limit} for 250~GHz.

The spectral-types of interest are single G, K and M giants. Unfortunately there were no single G giant detections reported 
in the two stellar samples, so we do not consider these objects further here. Because we are interested in the chromospheric emission, we need to avoid stars with significant ionized wind opacity, as is likely to occur
at lower gravity (higher luminosity), e.g., the eclipsing binary $\zeta$~Aur 
(K4~Ib + B5~V) \citep{harper_etal2005} and its spectral-type proxy $\lambda$~Vel (K4~Ib) \citep{carpenter_etal1999}, and earlier spectral types. We also exclude the hybrid chromosphere stars
$\gamma$~Aql (K3~II) and $\iota$~Aur (K3~II) because the hybrid $\alpha$~TrA (K3~II)
shows evidence for a warm ionized wind \citep{harper_etal2005}. We include $\alpha$~Boo
(K2~III) even though it is thought to have a 10,000~K partially ionized wind, 
 because the mass-loss rate 
$\dot{M}=2\times 10^{-10}\>{\rm M}_\odot{\rm yr}^{-1}$ \citep{drake1985} is small, although perhaps
not entirely negligible.  
For the later spectral-type semi-regular M giants
it is not clear when the energy and momentum deposited by 
small amplitude radial pulsations begins dominate the sources present in 
earlier spectral-type red giants and begin to fundamentally alter 
the chromospheric structure, see for example \citet{eaton_etal1990}.
We thus include R~Lyr (M5 III), while noting
that \citet*{luttermoser_etal1994} found it was not possible to construct a 
time-independent semi-empirical model for g~Her (M6~III).

For the sample, we have estimates and errors for $\phi$, $T_{\rm eff}$ and
$\hat{F}$. In Figure 3 we directly compare the model (with $G=1.08$) to
\citet{altenhoff_etal1994}. The model
is broadly consistent with the observed fluxes but because of the large uncertainties ($\ge 15$\%) of the
IRAM measurements it is not clear whether there is additional intrinsic scatter that is not
accounted for in the model. Combined Array for Research in Millimeter-wave Astronomy (CARMA) and ALMA observations with flux signal-to-noise ratios
of $\sim 20$ will be a defining test of our simple model of time-independent but 
inhomogeneous red giant chromospheres.

\subsection{Spectral Indices}

Spectral indices, $\alpha$ (where $F_\nu \propto \nu^\alpha$), are an empirical measure to help 
interpret multi-frequency radio observation.
For a constant temperature source with an angular diameter independent of frequency $\alpha=2$, reflecting the
frequency factor in the Planck function. Many $\alpha$'s have been obtained at quite different frequencies where the flux
can arise from different atmospheric layers (e.g., at cm-wavelengths 
these can reflect, say, the wind acceleration at 6~cm but the chromosphere
at 2~cm), which limits their utility. With the greater sensitivity of the Karl G. Jansky Very Large Array (JVLA) and ALMA,
a finer sweep of frequencies can be made and the resulting spectral energy distributions should provide more
powerful diagnostic information. 

Using the model above we see that the 100~GHz flux is predicted to come from the upper portion of the chromosphere: by 250~GHz the chromosphere
is becoming transparent; while at 900~GHz the chromosphere can be
become quite transparent, revealing the cold molecular layers above the photosphere, as seen in the CO 
fundamental bands \citep{wiedemann_etal1994}. For the sample in Table~2, 
we find spectral indices using ALMA Band 3 (100~GHz) and Band 9 (660~GHz) lie in the range 
$\sim 1.60 \pm 0.15$ indicating that the mean temperatures are
declining with increasing frequency.

\subsection{Variability}

The mm optical depths scale with chromospheric heating, so that observed changes in major cooling channel fluxes in the uv can reasonably be expected to induce corresponding
changes in $\Delta\tau$, i.e., $\delta{\cal F} \propto f\left(\delta\Delta\tau\right)$.
Few quantitative studies exist of the temporal chromospheric flux variability of red giants, although both $\alpha$~Boo and $\alpha$~Tau have been well observed with {\it IUE}. For $\alpha$~Tau, \citet{cuntz_etal1996} find a peak to peak Mg~II h \& k flux variations of $\pm 4$\% on day time scales, whereas
\citet*{robinson_etal1998} find a slightly larger $\sim 15$\% from three HST observations over seven years. 
\citet{mcclintock_etal1978} found little evidence for variability in $\alpha$~Boo's chromospheric emissions above {\it Copernicus's} instrumental noise: 25\% over tens of hours and 20\% on year time scales.
\cite{ayres_etal1995} reported a 6\% rms variation in the Mg~II h line observed by {\it IUE} over nearly a decade, although a smaller variability on day time scales was potentially masked by a long-term $\sim 10$\% secular trend.
\cite{brown_etal2008} find long term  Ca~II H \& K variations at the 10\% level
with a suggestion of a $>14$ year cycle. $\alpha$~Boo has a significant Mg~II h \& k emission above the {\it basal-flux} and it is possible that
the excess magnetic heating is responsible for changes in the chromosphere and mm-radio emission (see also discussion in \citealt{sennhauser2011}).
 \citet{judge_etal1993} have examined the chromospheric flux variations of
M giants, $\rho$~Per (M4~II, $\pm 10$\%), R~Lyr (M5~III, $\pm 4$\%), 
and g~Her (M6~III, $\pm 10$\%). These changes do not follow the variations 
in V magnitude. Perturbing the $\Delta\tau$'s in the analytic model by $\pm 15$\% imposes a 
fractional change in $\langle T_{\rm Br}\rangle$ (or 250~GHz flux) of $< 5$\%. 

On the basis of these studies we do not expect significant changes in the mm and sub-mm radio fluxes if the {\it chromospheric material is uniformly distributed}, such as may result from a high filling factor of the 
warm chromosphere, and/or spatial-averaged acoustic heating.

\subsection{Comparison with Recent Observations}

Red giants are feeble thermal radio emitters and obtaining accurate and precise fluxes is difficult.
Some more recent observations have been made of red giants by \citet{cohen_etal2005} 
(interferometric) and \citet{dehaes_etal2011} (single dish bolometer arrays). \citet{cohen_etal2005} have reported BIMA mm-fluxes for $\alpha$~Tau and $\alpha$~Boo
obtained at 217.8~GHz (1.38\,mm) and 108.4~GHz (2.77\,mm), and compared againt predictions based on
photospheric {\it Radiative Equilibrium (RE)} values. \citet{cohen_etal2005}
reported evidence for excess mm-emission from $\alpha$~Boo, which is consistent with the observed
presence chromospheric heating. Further, the BIMA fluxes imply little change in temperature with
frequency, whereas the present model reflects the decline of temperature as
the higher frequencies probe closer to the photosphere.
The 217.8~GHz flux for $\alpha$~Tau (when scaled) remains significantly lower than the IRAM 250~GHz flux.
A comparison of the BIMA, {\it RE} and present model is given in Table 3.

\begin{table}
\caption{Observed and predicted mm fluxes (mJy) for $\alpha$~Tau and $\alpha$~Boo. Observed values are from \citet{cohen_etal2005} and their predictions are based on radiative equilbrium models.}

\begin{tabular}{lll}
\hline  
                & $\alpha$~Boo (K2~III) &  $\alpha$~Tau (K5~III)\\ \hline
                & \multicolumn{2}{c}{217.8 GHz}   \\
Observed        & $83.50\pm 17.1$ & $25.78\pm 5.64$  \\ 
Cohen predicted & 33.43           & 32.38 \\
Present Model   & $\ge 52.8$      & 40.8  \\ 
\hline
                & \multicolumn{2}{c}{108.4 GHz}   \\
Observed        & $20.09\pm 0.69$ & $13.97\pm 1.46$  \\ 
Cohen Predicted & 7.94            & 7.80  \\
Present Model   & $\ge 17.7$      & 15.9  \\ 
\hline
\end{tabular}
\end{table}

\begin{table*}
\begin{minipage}{126mm}
\caption{Observed and predicted 1.2\,mm fluxes (mJy) for the standard stars of \citet{dehaes_etal2011}.
Note that fluxes given within `[]' were discounted because of weather conditions.}

\begin{tabular}{lllll}
\hline  
                       & MAMBO II        & SIMBA    & IRAM  & Predicted \\ \hline
Star                   & \multicolumn{2}{c}{Dehaes et al.} & Altenhoff et al. &      \\
$\alpha$~Boo (K2~III)  & [$20.8\pm 3.5$] & $105.6\pm 6.9$ & $78\pm 8$ & $\ge 57$  \\ 
$\beta$~And (M0~III)   & $23.5 \pm 2.7$  & $<40.1$        & $25\pm 4$ & 23  \\
$\alpha$~Cet (M2~III)  & [$23.6\pm 2.5$] & $58.3 \pm 4.1$ & $15\pm 3$ & 19  \\
$\beta$~Peg (M2.5~III) & $29.5\pm 3.2$   & $29.2 \pm 2.5$ & $23\pm 5$ & 32   \\
$\iota$~Aur (K3~II)    & $ 5.5 \pm 1.6$  & -              & $13\pm 3$ & $\ge 8$   \\
$\beta$~Umi (K4~III)   & $12.2 \pm 1.6$  & -              & $16\pm 4$ & 13   \\
$\gamma$~Dra (K5~III)  & $10.1 \pm 1.3$  & -              & $< 12$    & 12   \\
$\sigma$~Lib (M3~III)  & $12.1 \pm 2.0$  & -              & -         & 14   \\
\hline
\end{tabular}
\end{minipage}
\end{table*}

\citet{dehaes_etal2011} present 1.2\,mm SIMBA and MAMBO~II
observations for a sample of red giants. We compare these measurements to the
model in Table 4. In general the agreement is quite good, except for $\alpha$~Cet
(M2~III). The SIMBA flux of $58.3\pm 4.1$\,mJy is much greater than their MAMBO~II value of $23.6\pm 2.5$\,mJy,
which the authors discount owing to unstable weather conditions. The later, nevertheless, is much closer to the IRAM and model fluxes. The SIMBA flux for $\alpha$~Boo is close to the \citet{cohen_etal2005} value (when adjusted for the different frequencies), whereas our prediction is closer to IRAM.
These measurements were made at different epochs, however.

The limited data hint at variability at mm-wavelengths that is larger than
one might anticipate based on the small observed chromospheric variability
(peak-to-peak 15\% that implies 4\% in mm fluxes)

\section{Discussion}

The time-independent inhomogeneous model appears to be consistent with most existing 
mm-radio observations. However, because most measurements were made at a single frequency
and have sizeable flux uncertainties, these data are only just beginning to provide
insight into the chromospheres of inactive K and M giants. The characteristic temperatures
of stellar chromospheres: $T^{\rm uv}_{\rm Min}$, $T_{\rm CO}$, $T_{\rm Int}$, and $T_{\rm Chrom}$ all are
within 50\% of the mean, and the unfortunate reality is that there is limited thermal dynamic range in this problem. The errors on $\phi_\ast$ and $T_{\rm eff}$ are at the few percent level so to test the limits of this model will require multi-frequency observations with S/N $\sim 20$ and as good an absolute calibration as possible ($\sim 5$\%), both of which are challenging. Multiple observations at closely spaced frequencies are highly desirable, 
because each flux would have a contribution from overlapping spatial regions and provide a much better measure of the mean temperature gradient than spectral indices based on widely separated frequencies.
On this point we note that using $T_{\rm eff}$ and $\phi_\ast$ to estimate
the mm-radio flux in equation~(\ref{eq:convenient}) can give reasonable values even though it is a physically 
unsound model.

We have not explored the G giants, but the model could be extended to include a filling factor of hotter material 
not at the boundary layer temperature \citep{wiedemann_etal1994} by setting, for example, $T_{\rm CO}=T^{\rm uv}_{\rm Min}$ for that component.

\subsection{Variability}

Millimetre observations, such as those from JVLA Q-band, CARMA, and ALMA, have the potential to distinguish between the debated acoustic or magnetic origin of chromospheric basal-flux heating. Numerical simulations show that granule sizes for main-sequence and giant stars scale with the surface pressure scale-height 
\citep{freytag_etal2002, nordlund_dravins1990} and the grain size is $\sim 10H_{\rm pres}$. 
The number of granules on the stellar surface will scale as
\begin{equation}
N_{\rm gran} \simeq 4\times 10^6 \left({R_\odot\over{R_\star}}\right)^2\left({T_{{\rm eff},\odot}\over{T_{{\rm eff},\star}}}\right)^2 \left({M_\star\over{M_\odot}}\right)^2 
\end{equation}

which for $\alpha$~Tau gives $\sim 10,000$ granules on the visible disk. This scaling appears
consistent with findings from {\it Kepler} \citep{mathur_etal2011}.
If the horizontal coherence scale of the acoustic shock heating is of the order of 
the granule size 
\citep{loukitcheva_etal2008} then the
spatially unresolved radio flux will have contributions 
from $\sim 10,000$ phase-uncorrelated patches.
In the acoustic shocked heated interpretation, at any one moment there 
are likely to be several shocks passing through each patch, so one might 
expect that a basal-flux red giant to show quite small variability
at mm wavelengths in this scenario. Barring limb projection effects, the rms fractional fluctuations 
of the total intensity would be of order $1/\sqrt{10,000}=0.01$.
However, if observations show significant variability, then it is likely that
there is an additional, spatially more correlated source of chromospheric 
heating, i.e., magnetic active regions or 
possibly non-radial pulsations. 
To explore this aspect, it would be important to observe giant stars 
whose Mg~II h \& k or Ca~II H \& K fluxes are
as close to the basal limit as possible.

\subsection{Molecular Component}

In the Sun the temperature minimum appears to be set by the balance between mechanical heating and cooling by H$^-$. Above this temperature minimum there 
appears to be an unknown fraction, perhaps as much as 80\%, of the surface covered by CO clouds or columns \citep{ayres2002}. In the red giants the CO, which appears to be close to radiative equilibrium temperatures, appears to cover most of the stellar surfaces. 
g~Her (30~Her, M6~III, $\phi=19$\,mas, $T_{\rm eff}=3010$\,K) was not detected
in the \citet{altenhoff_etal1994} survey ($<6$\,mJy) and the
upper-limit to its brightness temperature is only 465~K. The present model predicts 28\,mJy and this 
is very discrepant. The star has been observed with both {\it IUE} \citep{judge_etal1993} and {\it HST} \citep{luttermoser_etal1994} and it shows uv chromospheric emission. 
g~Her and R~Lyr both have similar $\phi_\ast$ and $T_{\rm eff}$ and have similar 2\,cm radio VLA fluxes \citep{drake_etal1991}. In the model presented here we adopted $T_{\rm CO}=0.6T_{\rm eff}$, but if the low fluxes for
g~Her and R~Lyr are confirmed then the molecular component of the outer atmosphere
may be even cooler, perhaps dominated by an H$_2$O molsphere; and M5~III may mark the
onset of a cooler dominant component. However, even this would not seem enough to explain g~Her's low
1.2\,mm flux and multi-frequency observations are urgently needed. $\gamma$~Cru (M3~III) 
has been observed with the ATCA at 94~GHz . It has a large angular diameter 
($\phi_\ast\simeq 25$~mas, $T_{\rm eff}=3626$\,K; \citealt{ireland_etal2004,cohen_etal1996}) and the 
present model predicts $\sim 16$\,mJy but the observed flux is significantly less (M. Cohen, priv comm.).

It would seem that K and early M giants at the {\it basal-limit} might be suitable mm flux 
calibrators at the 5\% level. Random fluctuations in the brightness resulting from changes at the stellar 
photospheric granule scale should average down, and chromospheric heating fluctuations derived from optical and uv emission suggest small variations in the mm-fluxes. For later spectral-types, the uncertain influence of this variable chromospheric heating is further reduced.
That being said, the emission from the molecular material mixed in with warmer chromospheric plasma needs
to be calibrated in some way, because it is challenging to establish directly from theoretical models or simulations of stellar
atmospheres. 

The semi-analytic model developed here is a time-independent 1-D model with all the inherent assumptions. When the next 
generation of theoretical chromospheric heating models becomes available, one could analyse the thermodynamic structure and 
develop a new analytic representation. Until then, the present model, 
based on the interpretation of empirical evidence should be regarded as a starting point.
Even if real red giant chromospheres are highly time variable, the semi-empirical models upon which our
mm radio predictor is based, still represent a measure of the temperatures, electron densities, and ionization required 
to produce the uv and optical spectral diagnostics. One could imagine an actual atmosphere
as being a highly dynamic jumbled-up 1-D model, in which case the temperature rise in the warm chromosphere 
no longer is monotonic. The implication for the radio fluxes would be to reduce them slightly as cool material 
is now present at greater altitudes. This systematic bias is being investigated.

A test of chromospheric models for the largest angular diameter giants may be possible with ALMA, 
at frequencies above 250\,GHz, in its most extended (16\, km maximum baseline) configuration. The chromospheres 
will be partially resolved but they will provide constraints on the millimetre limb brightening or darkening.

\section{Conclusions}

We present a semi-analytic 'toy' chromospheric model to examine millimetre thermal continuum 
radio fluxes of non-dusty and non-pulsating red giants. Equation (\ref{eq:tbr}) shows, in a clear fashion, the contributions to the brightness temperature from the cool molecular layers 
at the base, and from the chromosphere itself, through weighting factors obtained from analytic solutions to the radiative transfer equation. At 3~mm, the chromospheres are semi-opaque; while at 1.2\,mm the
chromosphere becomes optically thin and the cool molecular material becomes visible. 
The model predicts a {\it basal} level of radio emission that depends on the temperature of 
the boundary layers. Studies at (sub-) mm will provide new insights into these especially enigmatic molecular regions. 
The model also permits an analysis of radio flux variability in response to changes in 
chromospheric heating as observed in Mg~II h \& k or Ca~II H \& K. If the chromospheric heating rate is
not known for a given red giant, the model still can predict the basal mm radio fluxes. Significant variations in radio emission for 
{\it basal-flux} red giants, if detected, would indicate a source of structured  large-scale heating, possibly magnetic in nature.
The physical arrangement, e.g., columns and sheets, of the cool and warm chromospheric plasma can 
be constrained by the mm-radio limb-brightening of red giants that can be spatially resolved when ALMA 
reaches its maximum baselines. A good test of this semi-analytic model would be to compare it against red giants with 
independently determined $T_{\rm eff}$, $\phi_\ast$, and $T_{\rm CO}$ and that can be monitored for changes in chromospheric heating.

In summary, the present analytic model is consistent with most existing radio measurements and also highlights 
some possible anomalies. The sensitivity of mm-fluxes to changes in chromospheric heating can be evaluated and 
spectral indices are predicted for the existing thermal gradients. When a sample of multiwavelength (sub-) mm-radio 
observations with good signal-to-noise becomes available these can be used to test the overall validity of the model.
If the model passes this test, i.e., if there are not significant star to star discrepancies, the observations can then 
be used to refine and test the mean thermal gradient in the stellar chromospheres. The spectral shapes could 
be used to better describe the outward temperature gradient, or alternately they could invalidate the present 
model completely if there is no outward temperature increase. 
More importantly, these important new radio observations can test new 3-D theoretical heating models 
when they become available.

There appear to be a few stars with significant 1.2\,mm flux differences
between widely separated epochs and different observing facilities that appear at first sight to be larger than
one would expect based on observed changes in chromospheric heating. These would
suggest significant magnetic contributions for stars at twice the basal uv-flux
levels. We also have identified a few of late M giants which, while at the limit of applicability of the model, 
seem to have very low observed fluxes. However, we must await 
accurate new observations of these stars before we can draw any significant conclusions.

While optical angular diameters and effective temperatures for many giants are known to a precision and 
accuracy of a few percent, we have shown that the physical dynamic range of $T_{\rm Br}$ for the mm-radio problem is limited. 
Future progress will require high S/N multi-wavelength (sub-) mm radio fluxes with as small absolute 
flux calibration errors as possible. 

\section*{Acknowledgements}

We would like to thank Drs. D. Luttermoser, A. McMurry, and K. Eriksson for providing
us with their semi-empirical models, and Dr. K. Carpenter for providing us with Mg~II h \& k fluxes
for $\alpha$~Tau. This publication
has emanated from research conducted with the
financial support of Science Foundation Ireland
under Grant Number SFI11/RFP.1/AST/3064,
and a grant from Trinity College Dublin.

\bibliography{harper}

\begin{thebibliography}{69}
\expandafter\ifx\csname natexlab\endcsname\relax\def\natexlab#1{#1}\fi

\bibitem[{{Altenhoff} {et~al.}(1960){Altenhoff}, {Mezger}, {Wendker}, \&
  {Westerhout}}]{altenhoff_etal1960}
{Altenhoff} W.~J., {Mezger} P.~G., {Wendker} H.~J., {Westerhout} G., 1960,
  Ver\"off. Sternw\"arte, Bonn, 59, 48

\bibitem[{{Altenhoff} {et~al.}(1994){Altenhoff}, {Thum}, \&
  {Wendker}}]{altenhoff_etal1994}
{Altenhoff} W.~J., {Thum} C., {Wendker} H.~J., 1994, A\&A, 281, 161

\bibitem[{{Ayres}(1979)}]{ayres1979}
{Ayres} T.~R., 1979, ApJ, 228, 509

\bibitem[{{Ayres}(2002)}]{ayres2002}
---, 2002, ApJ, 575, 1104

\bibitem[{{Ayres} {et~al.}(1995){Ayres}, {Fleming}, {Simon}, {Haisch}, {Brown},
  {Lenz}, {Wamsteker}, {de Martino}, {Gonzalez}, {Bonnell}, {Mas-Hesse},
  {Rosso}, {Schmitt}, {Truemper}, {Voges}, {Pye}, {Dempsey}, {Linsky},
  {Guinan}, {Harper}, {Jordan}, {Montesinos}, {Pagano}, \&
  {Rodono}}]{ayres_etal1995}
{Ayres} T.~R., {Fleming} T.~A., {Simon} T., {Haisch} B.~M., {Brown} A., {Lenz}
  D., {Wamsteker} W., {de Martino} D., {Gonzalez} C., {Bonnell} J., {Mas-Hesse}
  J.~M., {Rosso} C., {Schmitt} J.~H.~M.~M., {Truemper} J., {Voges} W., {Pye}
  J., {Dempsey} R.~C., {Linsky} J.~L., {Guinan} E.~F., {Harper} G.~M., {Jordan}
  C., {Montesinos} B.~M., {Pagano} I., {Rodono} M., 1995, ApJS, 96, 223

\bibitem[{{Ayres} \& {Linsky}(1975)}]{ayres_linsky1975}
{Ayres} T.~R., {Linsky} J.~L., 1975, ApJ, 200, 660

\bibitem[{{Ayres} {et~al.}(1974){Ayres}, {Linsky}, \& {Shine}}]{ayres_etal1974}
{Ayres} T.~R., {Linsky} J.~L., {Shine} R.~A., 1974, ApJ, 192, 93

\bibitem[{{Ayres} \& {Testerman}(1981)}]{ayres_testerman1981}
{Ayres} T.~R., {Testerman} L., 1981, ApJ, 245, 1124

\bibitem[{{Baines} {et~al.}(2010){Baines}, {D{\"o}llinger}, {Cusano},
  {Guenther}, {Hatzes}, {McAlister}, {ten Brummelaar}, {Turner}, {Sturmann},
  {Sturmann}, {Goldfinger}, {Farrington}, \& {Ridgway}}]{baines_etal2010}
{Baines} E.~K., {D{\"o}llinger} M.~P., {Cusano} F., {Guenther} E.~W., {Hatzes}
  A.~P., {McAlister} H.~A., {ten Brummelaar} T.~A., {Turner} N.~H., {Sturmann}
  J., {Sturmann} L., {Goldfinger} P.~J., {Farrington} C.~D., {Ridgway} S.~T.,
  2010, ApJ, 710, 1365

\bibitem[{{Berio} {et~al.}(2011){Berio}, {Merle}, {Th{\'e}venin}, {Bonneau},
  {Mourard}, {Chesneau}, {Delaa}, {Ligi}, {Nardetto}, {Perraut}, {Pichon},
  {Stee}, {Tallon-Bosc}, {Clausse}, {Spang}, {McAlister}, {Ten Brummelaar},
  {Sturmann}, {Sturmann}, {Turner}, {Farrington}, \&
  {Goldfinger}}]{berio_etal2011}
{Berio} P., {Merle} T., {Th{\'e}venin} F., {Bonneau} D., {Mourard} D.,
  {Chesneau} O., {Delaa} O., {Ligi} R., {Nardetto} N., {Perraut} K., {Pichon}
  B., {Stee} P., {Tallon-Bosc} I., {Clausse} J.~M., {Spang} A., {McAlister} H.,
  {Ten Brummelaar} T., {Sturmann} J., {Sturmann} L., {Turner} N., {Farrington}
  C., {Goldfinger} P.~J., 2011, A\&A, 535, A59

\bibitem[{{Brown} {et~al.}(2008){Brown}, {Gray}, \&
  {Baliunas}}]{brown_etal2008}
{Brown} K.~I.~T., {Gray} D.~F., {Baliunas} S.~L., 2008, ApJ, 679, 1531

\bibitem[{{Buchholz} {et~al.}(1998){Buchholz}, {Ulmschneider}, \&
  {Cuntz}}]{buchholz_etal1998}
{Buchholz} B., {Ulmschneider} P., {Cuntz} M., 1998, ApJ, 494, 700

\bibitem[{{Butler} \& {Wooten}(1999)}]{butler_wooten1999}
{Butler} B., {Wooten} A., 1999, ALMA Memo., 276, 1

\bibitem[{{Byrne} {et~al.}(1988){Byrne}, {Murphy}, {Dufton}, {Kingston}, \&
  {Lennon}}]{byrne_etal1988}
{Byrne} P.~B., {Murphy} H.~M., {Dufton} P.~L., {Kingston} A.~E., {Lennon}
  D.~J., 1988, A\&A, 197, 205

\bibitem[{{Carlsson} \& {Stein}(1995)}]{carlsson_stein1995}
{Carlsson} M., {Stein} R.~F., 1995, ApJ, 440, L29

\bibitem[{{Carpenter} {et~al.}(1999){Carpenter}, {Robinson}, {Harper},
  {Bennett}, {Brown}, \& {Mullan}}]{carpenter_etal1999}
{Carpenter} K.~G., {Robinson} R.~D., {Harper} G.~M., {Bennett} P.~D., {Brown}
  A., {Mullan} D.~J., 1999, ApJ, 521, 382

\bibitem[{{Cohen} {et~al.}(2005){Cohen}, {Carbon}, {Welch}, {Lim}, {Schulz},
  {McMurry}, {Forster}, \& {Goorvitch}}]{cohen_etal2005}
{Cohen} M., {Carbon} D.~F., {Welch} W.~J., {Lim} T., {Schulz} B., {McMurry}
  A.~D., {Forster} J.~R., {Goorvitch} D., 2005, AJ, 129, 2836

\bibitem[{{Cohen} {et~al.}(1996){Cohen}, {Witteborn}, {Carbon}, {Davies},
  {Wooden}, \& {Bregman}}]{cohen_etal1996}
{Cohen} M., {Witteborn} F.~C., {Carbon} D.~F., {Davies} J.~K., {Wooden} D.~H.,
  {Bregman} J.~D., 1996, AJ, 112, 2274

\bibitem[{{Cuntz} {et~al.}(1996){Cuntz}, {Deeney}, {Brown}, \&
  {Stencel}}]{cuntz_etal1996}
{Cuntz} M., {Deeney} B.~D., {Brown} A., {Stencel} R.~E., 1996, ApJ, 464, 426

\bibitem[{{Decin} {et~al.}(2003){Decin}, {Vandenbussche}, {Waelkens}, {Decin},
  {Eriksson}, {Gustafsson}, {Plez}, \& {Sauval}}]{decin_etal2003}
{Decin} L., {Vandenbussche} B., {Waelkens} C., {Decin} G., {Eriksson} K.,
  {Gustafsson} B., {Plez} B., {Sauval} A.~J., 2003, A\&A, 400, 709

\bibitem[{{Dehaes} {et~al.}(2011){Dehaes}, {Bauwens}, {Decin}, {Eriksson},
  {Raskin}, {Butler}, {Dowell}, {Ali}, \& {Blommaert}}]{dehaes_etal2011}
{Dehaes} S., {Bauwens} E., {Decin} L., {Eriksson} K., {Raskin} G., {Butler} B.,
  {Dowell} C.~D., {Ali} B., {Blommaert} J.~A.~D.~L., 2011, A\&A

\bibitem[{{Drake}(1985)}]{drake1985}
{Drake} S.~A., 1985, in NATO ASIC Proc. 152: Progress in Stellar Spectral Line
  Formation Theory, {J.~E.~Beckman \& L.~Crivellari}, ed., pp. 351--357

\bibitem[{{Drake} {et~al.}(1991){Drake}, {Linsky}, {Judge}, \&
  {Elitzur}}]{drake_etal1991}
{Drake} S.~A., {Linsky} J.~L., {Judge} P.~G., {Elitzur} M., 1991, AJ, 101, 230

\bibitem[{{Eaton}(1993)}]{eaton1993}
{Eaton} J.~A., 1993, ApJ, 404, 305

\bibitem[{{Eaton} {et~al.}(1990){Eaton}, {Johnson}, \&
  {Cadmus}}]{eaton_etal1990}
{Eaton} J.~A., {Johnson} H.~R., {Cadmus} Jr. R.~R., 1990, ApJ, 364, 259

\bibitem[{{Eriksson} {et~al.}(1983){Eriksson}, {Linsky}, \&
  {Simon}}]{eriksson_etal1983}
{Eriksson} K., {Linsky} J.~L., {Simon} T., 1983, ApJ, 272, 665

\bibitem[{{Fontenla} {et~al.}(1990){Fontenla}, {Avrett}, \& {Loeser}}]{fal1990}
{Fontenla} J.~M., {Avrett} E.~H., {Loeser} R., 1990, ApJ, 355, 700

\bibitem[{{Fontenla} {et~al.}(2009){Fontenla}, {Curdt}, {Haberreiter},
  {Harder}, \& {Tian}}]{fontenla2009}
{Fontenla} J.~M., {Curdt} W., {Haberreiter} M., {Harder} J., {Tian} H., 2009,
  ApJ, 707, 482

\bibitem[{{Freytag} {et~al.}(2002){Freytag}, {Steffen}, \&
  {Dorch}}]{freytag_etal2002}
{Freytag} B., {Steffen} M., {Dorch} B., 2002, Astronomische Nachrichten, 323,
  213

\bibitem[{{Harper}(1992)}]{harper1992}
{Harper} G.~M., 1992, MNRAS, 256, 37

\bibitem[{{Harper} {et~al.}(2005){Harper}, {Brown}, {Bennett}, {Baade},
  {Walder}, \& {Hummel}}]{harper_etal2005}
{Harper} G.~M., {Brown} A., {Bennett} P.~D., {Baade} R., {Walder} R., {Hummel}
  C.~A., 2005, AJ, 129, 1018

\bibitem[{{Harper} {et~al.}(2001){Harper}, {Brown}, \& {Lim}}]{harper_etal2001}
{Harper} G.~M., {Brown} A., {Lim} J., 2001, ApJ, 551, 1073

\bibitem[{{Hummer}(1988)}]{hummer1988}
{Hummer} D.~G., 1988, ApJ, 327, 477

\bibitem[{{Ireland} {et~al.}(2004){Ireland}, {Tuthill}, {Bedding}, {Robertson},
  \& {Jacob}}]{ireland_etal2004}
{Ireland} M.~J., {Tuthill} P.~G., {Bedding} T.~R., {Robertson} J.~G., {Jacob}
  A.~P., 2004, MNRAS, 350, 365

\bibitem[{{Judge} \& {Carpenter}(1998)}]{judge_carpenter1998}
{Judge} P.~G., {Carpenter} K.~G., 1998, ApJ, 494, 828

\bibitem[{{Judge} {et~al.}(1993){Judge}, {Luttermoser}, {Neff}, {Cuntz}, \&
  {Stencel}}]{judge_etal1993}
{Judge} P.~G., {Luttermoser} D.~G., {Neff} D.~H., {Cuntz} M., {Stencel} R.~E.,
  1993, AJ, 105, 1973

\bibitem[{{Judge} \& {Stencel}(1991)}]{judge_stencel1991}
{Judge} P.~G., {Stencel} R.~E., 1991, ApJ, 371, 357

\bibitem[{{Kalkofen} {et~al.}(1999){Kalkofen}, {Ulmschneider}, \&
  {Avrett}}]{kalkofen_etal1999}
{Kalkofen} W., {Ulmschneider} P., {Avrett} E.~H., 1999, ApJ, 521, L141

\bibitem[{{Kelch} {et~al.}(1978){Kelch}, {Chang}, {Furenlid}, {Linsky},
  {Basri}, {Chiu}, \& {Maran}}]{kelch_etal1978}
{Kelch} W.~L., {Chang} S.-H., {Furenlid} I., {Linsky} J.~L., {Basri} G.~S.,
  {Chiu} H.-Y., {Maran} S.~P., 1978, ApJ, 220, 962

\bibitem[{{Leenaarts} {et~al.}(2011){Leenaarts}, {Carlsson}, {Hansteen}, \&
  {Gudiksen}}]{leenaarts_etal2011}
{Leenaarts} J., {Carlsson} M., {Hansteen} V., {Gudiksen} B.~V., 2011, A\&A,
  530, A124

\bibitem[{{Linsky} \& {Avrett}(1970)}]{linsky_avrett1970}
{Linsky} J.~L., {Avrett} E.~H., 1970, PASP, 82, 169

\bibitem[{{Linsky} \& {Ayres}(1978)}]{linsky_ayres1978}
{Linsky} J.~L., {Ayres} T.~R., 1978, ApJ, 220, 619

\bibitem[{{Loukitcheva} {et~al.}(2004){Loukitcheva}, {Solanki}, {Carlsson}, \&
  {Stein}}]{loukitcheva_etal2004}
{Loukitcheva} M., {Solanki} S.~K., {Carlsson} M., {Stein} R.~F., 2004, A\&A,
  419, 747

\bibitem[{{Loukitcheva} {et~al.}(2008){Loukitcheva}, {Solanki}, \&
  {White}}]{loukitcheva_etal2008}
{Loukitcheva} M.~A., {Solanki} S.~K., {White} S., 2008, Astrophys Space Sci,
  313, 197

\bibitem[{{Luttermoser} {et~al.}(1994){Luttermoser}, {Johnson}, \&
  {Eaton}}]{luttermoser_etal1994}
{Luttermoser} D.~G., {Johnson} H.~R., {Eaton} J., 1994, ApJ, 422, 351

\bibitem[{{Mart{\'{\i}}nez} {et~al.}(2011){Mart{\'{\i}}nez}, {Schr{\"o}der}, \&
  {Cuntz}}]{martinez_etal2011}
{Mart{\'{\i}}nez} M.~I.~P., {Schr{\"o}der} K.-P., {Cuntz} M., 2011, MNRAS, 348

\bibitem[{{Mathur} {et~al.}(2011){Mathur}, {Hekker}, {Trampedach}, {Ballot},
  {Kallinger}, {Buzasi}, {Garc{\'{\i}}a}, {Huber}, {Jim{\'e}nez}, {Mosser},
  {Bedding}, {Elsworth}, {R{\'e}gulo}, {Stello}, {Chaplin}, {De Ridder},
  {Hale}, {Kinemuchi}, {Kjeldsen}, {Mullally}, \& {Thompson}}]{mathur_etal2011}
{Mathur} S., {Hekker} S., {Trampedach} R., {Ballot} J., {Kallinger} T.,
  {Buzasi} D., {Garc{\'{\i}}a} R.~A., {Huber} D., {Jim{\'e}nez} A., {Mosser}
  B., {Bedding} T.~R., {Elsworth} Y., {R{\'e}gulo} C., {Stello} D., {Chaplin}
  W.~J., {De Ridder} J., {Hale} S.~J., {Kinemuchi} K., {Kjeldsen} H.,
  {Mullally} F., {Thompson} S.~E., 2011, ApJ, 741, 119

\bibitem[{{McClintock} {et~al.}(1978){McClintock}, {Moos}, {Henry}, {Linsky},
  \& {Barker}}]{mcclintock_etal1978}
{McClintock} W., {Moos} H.~W., {Henry} R.~C., {Linsky} J.~L., {Barker} E.~S.,
  1978, ApJS, 37, 223

\bibitem[{{McMurry}(1999)}]{mcmurry1999}
{McMurry} A.~D., 1999, MNRAS, 302, 37

\bibitem[{{Mozurkewich} {et~al.}(2003){Mozurkewich}, {Armstrong}, {Hindsley},
  {Quirrenbach}, {Hummel}, {Hutter}, {Johnston}, {Hajian}, {Elias}, {Buscher},
  \& {Simon}}]{mozurkewich_etal2003}
{Mozurkewich} D., {Armstrong} J.~T., {Hindsley} R.~B., {Quirrenbach} A.,
  {Hummel} C.~A., {Hutter} D.~J., {Johnston} K.~J., {Hajian} A.~R., {Elias} II
  N.~M., {Buscher} D.~F., {Simon} R.~S., 2003, AJ, 126, 2502

\bibitem[{{Nordlund} \& {Dravins}(1990)}]{nordlund_dravins1990}
{Nordlund} A., {Dravins} D., 1990, A\&A, 228, 155

\bibitem[{{Ohnaka} {et~al.}(2012){Ohnaka}, {Hofmann}, {Schertl}, {Weigelt},
  {Malbet}, {Massi}, {Meilland}, \& {Stee}}]{ohnaka_etal2012}
{Ohnaka} K., {Hofmann} K.-H., {Schertl} D., {Weigelt} G., {Malbet} F., {Massi}
  F., {Meilland} A., {Stee} P., 2012, A\&A, 537, A53

\bibitem[{{Olson} \& {Kunasz}(1987)}]{olson_kunasz1987}
{Olson} G.~L., {Kunasz} P.~B., 1987, JQSRT, 38, 325

\bibitem[{{Perrin} {et~al.}(2004){Perrin}, {Ridgway}, {Mennesson}, {Cotton},
  {Woillez}, {Verhoelst}, {Schuller}, {Coud{\'e} du Foresto}, {Traub},
  {Millan-Gabet}, \& {Lacasse}}]{perrin_etal2004}
{Perrin} G., {Ridgway} S.~T., {Mennesson} B., {Cotton} W.~D., {Woillez} J.,
  {Verhoelst} T., {Schuller} P., {Coud{\'e} du Foresto} V., {Traub} W.~A.,
  {Millan-Gabet} R., {Lacasse} M.~G., 2004, A\&A, 426, 279

\bibitem[{{Quirrenbach} {et~al.}(1993){Quirrenbach}, {Mozurkewich},
  {Armstrong}, {Buscher}, \& {Hummel}}]{quirrenbach_etal1993}
{Quirrenbach} A., {Mozurkewich} D., {Armstrong} J.~T., {Buscher} D.~F.,
  {Hummel} C.~A., 1993, ApJ, 406, 215

\bibitem[{{Robinson} {et~al.}(1998){Robinson}, {Carpenter}, \&
  {Brown}}]{robinson_etal1998}
{Robinson} R.~D., {Carpenter} K.~G., {Brown} A., 1998, ApJ, 503, 396

\bibitem[{{Rowe}(1992)}]{rowe1992}
{Rowe} A.~K., 1992, in Astronomical Society of the Pacific Conference Series,
  Vol.~26, Cool Stars, Stellar Systems, and the Sun, {Giampapa} M.~S.,
  {Bookbinder} J.~A., eds., p. 561

\bibitem[{{Rybicki} \& {Lightman}(2004)}]{rybicki_lightman2004}
{Rybicki} G.~B., {Lightman} A.~P., 2004, {Radiative Processes in Astrophysics,
  Wiley-VCH Verlag GmbH \& Co. KGaA}

\bibitem[{{Schrijver} \& {Zwaan}(2000)}]{schrijver_zwaan2000}
{Schrijver} C.~J., {Zwaan} C., 2000, {Solar and Stellar Magnetic Activity,
  Cambridge Astrophysics Series 34, Cambridge University Press}

\bibitem[{{Sennhauser} \& {Berdyugina}(2011)}]{sennhauser2011}
{Sennhauser} C., {Berdyugina} S.~V., 2011, A\&A, 529, A100

\bibitem[{{Tsuji}(2008)}]{tsuji2008}
{Tsuji} T., 2008, A\&A, 489, 1271

\bibitem[{{Tsuji} {et~al.}(1997){Tsuji}, {Ohnaka}, {Aoki}, \&
  {Yamamura}}]{tsuji_etal1997}
{Tsuji} T., {Ohnaka} K., {Aoki} W., {Yamamura} I., 1997, A\&A, 320, L1

\bibitem[{{Uitenbroek}(1989)}]{uitenbroek1989}
{Uitenbroek} H., 1989, A\&A, 213, 360

\bibitem[{{Ulmschneider}(1991)}]{ulmschneider1991}
{Ulmschneider} P., 1991, in Mechanisms of Chromospheric and Coronal Heating,
  {P.~Ulmschneider, E.~R.~Priest, \& R.~Rosner}, ed., pp. 328--+

\bibitem[{{Vecchio} {et~al.}(2009){Vecchio}, {Cauzzi}, \&
  {Reardon}}]{vecchio_etal2009}
{Vecchio} A., {Cauzzi} G., {Reardon} K.~P., 2009, A\&A, 494, 269

\bibitem[{{Vernazza} {et~al.}(1976){Vernazza}, {Avrett}, \& {Loeser}}]{val1976}
{Vernazza} J.~E., {Avrett} E.~H., {Loeser} R., 1976, ApJS, 30, 1

\bibitem[{{Vernazza} {et~al.}(1981){Vernazza}, {Avrett}, \& {Loeser}}]{val1981}
---, 1981, ApJS, 45, 635

\bibitem[{{Wedemeyer-B{\"o}hm} {et~al.}(2007){Wedemeyer-B{\"o}hm}, {Ludwig},
  {Steffen}, {Leenaarts}, \& {Freytag}}]{wedemeyerbohm_etal2007}
{Wedemeyer-B{\"o}hm} S., {Ludwig} H.~G., {Steffen} M., {Leenaarts} J.,
  {Freytag} B., 2007, A\&A, 471, 977

\bibitem[{{Wiedemann} {et~al.}(1994){Wiedemann}, {Ayres}, {Jennings}, \&
  {Saar}}]{wiedemann_etal1994}
{Wiedemann} G., {Ayres} T.~R., {Jennings} D.~E., {Saar} S.~H., 1994, ApJ, 423,
  806

\end{thebibliography}

\bsp

\label{lastpage}

\end{document}